\documentclass[prl,twocolumn,showpacs]{revtex4}

\usepackage{graphicx}
\usepackage{dcolumn}
\usepackage{bm}
\usepackage{amsmath}
  
\begin{document}

\title{Pressure-Induced Anomalous Phase Transitions and 
Colossal Enhancement of Piezoelectricity in PbTiO$_3$}

\author{Zhigang Wu}
\affiliation{Geophysical Laboratory, Carnegie Institution of Washington, Washington, DC 20015, USA}
\author{Ronald E. Cohen}
\affiliation{Geophysical Laboratory, Carnegie Institution of Washington, Washington, DC 20015, USA}

\date{\today}

\begin{abstract}
We find an unexpected tetragonal-to-monoclinic-to-rhombohedral-to-cubic phase transition sequence 
induced by pressure, and a morphotropic phase boundary in a pure compound using first-principles 
calculations. Huge dielectric and piezoelectric coupling constants occur in the transition 
regions, comparable to those observed in the new complex single-crystal solid-solution 
piezoelectrics such as Pb(Mg$_{1/3}$Nb$_{2/3}$)O$_{3}$-PbTiO$_{3}$, which 
are expected to revolutionize electromechanical applications. Our results show that 
morphotropic phase boundaries and giant piezoelectric effects do not require intrinsic disorder,
and open the possibility of studying this effect in simple systems.

\end{abstract}


\maketitle

The classic ferroelectric PbTiO$_3$ has been known to have a single ferroelectric tetragonal (T) 
to paraelectric cubic phase transition with increased temperature \cite{ref2} or pressure \cite{ref3}
since its discovery \cite{ref9} in 1950. It seemed unlikely to discover any new ferroelectric
transitions in PbTiO$_3$, but we predict new ferroelectric phases under pressure.
Piezoelectrics convert electric energy to mechanical energy, and vice versa. They are widely 
used in medical imaging, acoustic sensors and transducers, actuators, {\it etc} \cite{ref10,ref11}.
PbTiO$_3$ has been extensively studied \cite{ref1,ref2,ref3,ref4,ref5,ref6,ref7,ref8} to understand 
the electronic origin of ferroelectricity. The Pb-O and Ti-O hybridization in PbTiO$_3$ weakens 
the short-range repulsions and gives rise to the ferroelectric distortion \cite{ref4}. It has been 
found that pressure suppresses ferroelectricity since compression favours short-range repulsion, 
but there has been no previous work on the effect of pressure on piezoelectricity.  

PbTiO$_3$ has a high $c$-axis strain of 6.5\% (7.1\%) at room (low) temperature. Under ambient 
pressure it undergoes a first-order phase transition at $T_{\rm c} = 763$ K, and this transition is 
regarded as a typical displacive transition, which is associated with soft-modes \cite{ref2}. 
The displacive phase transition temperature $T_{\rm c}$ decreases under hydrostatic pressure. 
A Raman study of PbTiO$_3$ shows that $T_{\rm c}$ reduces to 300 K and the phase transition 
becomes second-order at $P = 12.1$ GPa \cite{ref3}. It indicates a tricritical point in the 
phase diagram, where the first-order phase transition changes to second-order, and it has 
been identified at $P = 1.75$ GPa, $T = 649$ K \cite{ref12}. Increasing 
pressure further reduces $T_{\rm c}$. 

PbTiO$_3$ is an end member of 
Pb(Mg$_{1/3}$Nb$_{2/3}$)O$_{3}$-PbTiO$_{3}$ (PMN-PT) and 
Pb(Zn$_{1/3}$Nb$_{2/3}$)O$_{3}$-PbTiO$_{3}$ (PZN-PT), which have piezoelectric
coefficients an order of magnitude larger than those of conventional ferroelectric simple compounds 
\cite{ref13}. It is also an end member of PbZrO$_{3}$-PbTiO$_3$ (PZT), the most widely used current 
piezoelectric, which is ubiquitous in modern technology. These materials have a common feature, namely 
the morphotropic phase boundary (MPB), and they have optimum piezoelectric efficiency near the MPB. 
Polarization rotation \cite{ref14,ref15} is believed to play an important role for this extraordinary 
property. Close to the MPB, the energy surface for polarization is very flat so that for example a 
polarization along [111] direction in the rhombohedral (R) phase can be easily rotated 
toward the tetragonal polarization direction by applying 
an electric field along [001] direction, and the R phase is transformed to the T phase via the 
intermediate phase(s). This mechanism has been elucidated by theoretical calculations on 
BaTiO$_{3}$ \cite{ref14} and PZT \cite{ref16,ref17}, and 
by experimental findings of low symmetry phases \cite{ref18,ref19,ref20}. The giant 
piezoelectric effects always do not occur along the spontaneous polarization 
direction \cite{ref13,ref15}). At ambient pressure PbTiO$_3$ has no such MPB because it has a rather 
stiff energy surface near the T phase and it has no R phase. Under high pressures the energy (enthalpy) 
surface is expected to be softer, and an MPB could arise. 

In this Letter we address both issues of phase transitions and piezoelectricity of PbTiO$_3$ under
hydrostatic pressures. We perform total energy as well as linear response computations, which
have been proved to be highly reliable for ground state properties. We find anomalous phase
transitions and giant enhancement of dielectricity and piezoelectricity near the phase transition 
regions induced by pressure.

We have carried out zero temperature {\it ab\ initio} computations based on density functional 
theory (DFT) within the local density approximation (LDA). We used the pseudopotential 
planewave method implemented in the ABINIT package \cite{ref21}. The planewave energy cutoff is 60 
Hartree, and the {\bf k}-point mesh for Brillouin Zone integration is of $6 \times 6 \times 6$. 
We used the OPIUM program \cite{ref22} to generate norm-conserving pseudopotentials, which were 
rigorously tested against the full-potential linearized augmented planewave (LAPW) method \cite{ref23}. 
We included semi-core states of Pb 5$d^{10}$, Ti 3$s^{2}$3$p^{6}$4$d^{2}$, and O 2$s^{2}$ 
in valence states. We chose the LDA instead of the generalized gradient approximation (GGA) because 
the GGA catastrophically overestimates both equilibrium volume and strain for tetragonal 
PbTiO$_3$ \cite{ref24}. On the other hand, the LDA moderately underestimates the volume 
(60.38 \AA$^{3}$) and strain (4.6\%), and the experimental $V_{0}$ 
corresponds to a negative pressure $P_{0} = -2.2$ GPa.
 
\begin{table}[tbp]
\caption{Elastic ($c_{\mu\nu}$) and piezoelectric ($e_{i\nu}$) constants of tetragonal $P4mm$
PbTiO$_3$ with lattice constants $a = 3.902$ \AA, $c = 4.155$ \AA. $c_{\mu\nu}$ and $e_{i\nu}$ 
are in GPa and C/m$^{2}$, respectively. Here FS refers to the finite strain method.
\label{tab1}}

\begin{ruledtabular}
\begin{tabular}{l|ccccccccc} 
 Methods & $c_{11}$ & $c_{12}$ & $c_{13}$ & $c_{33}$ & $c_{44}$ &  $c_{66}$ 
         & $e_{31}$ & $e_{33}$ & $e_{15}$ \\ 
\hline
 DFPT    & 230      & 96.2     & 65.2     & 41.9     & 46.6     & 98.8      
         & 2.06     & 4.41     & 6.63     \\
 FS      & 229      & 95.6     & 64.3     & 41.2     & 47.2     & 98.6      
         & 2.07     & 4.48     & 6.66     \\
\end{tabular}
\end{ruledtabular}
\end{table}

To calculate the dielectric susceptibility $\chi_{ij}$, elastic constants $c_{\mu\nu}$, and 
piezoelectric stress constants $e_{i\nu}$ (Here Latin indexes run from 1 to 3, and Greek indexes 
from 1 to 6), we used the density functional perturbation theory (DFPT) \cite{ref25} of the 
linear response of strain type perturbations \cite{ref26}. The DFPT is based on the systematic 
expansion of the variation expression of the DFT total energy in powers of parameters, such as 
atomic coordinates, macroscopic strain and electric field. Tensors $\chi$ and $c$ involve second 
derivatives of total energy with respect to electric field and strain respectively, and tensor 
$e$ is the mixed second derivative of total energy with respect to strain and electric field. Here 
$c_{\mu\nu} \equiv \frac{\partial \sigma_{\mu}}{\partial \eta_{\nu}}$ 
with $\sigma$ stress and $\eta$ strain. The DFPT calculates 
$C_{\mu\nu} = \frac{\partial^{2}E}{\partial \eta_{\mu}\partial \eta_{\nu}}$
with $E$ the total energy. Under non-zero stress (e.g. pressure), 
$c_{ijkl} = C_{ijkl} - \delta_{ij}\sigma_{kl}$. 
The elastic compliance tensor $s$ is the reciprocal of tensor $c$, and the piezoelectric 
strain coefficients 
$d_{i\nu} = \sum^{6}_{\mu=1} e_{i\mu} s_{\mu\nu}$. 
We have examined the accuracy of $c_{\mu\nu}$ 
and $e_{i\nu}$ obtained from the DFPT by comparing with the finite strain method. As summarized in 
Table \ref{tab1}, excellent agreement between these two methods is achieved. 

\begin{figure}[tbp]
\begin{center}
\includegraphics[width=\columnwidth]{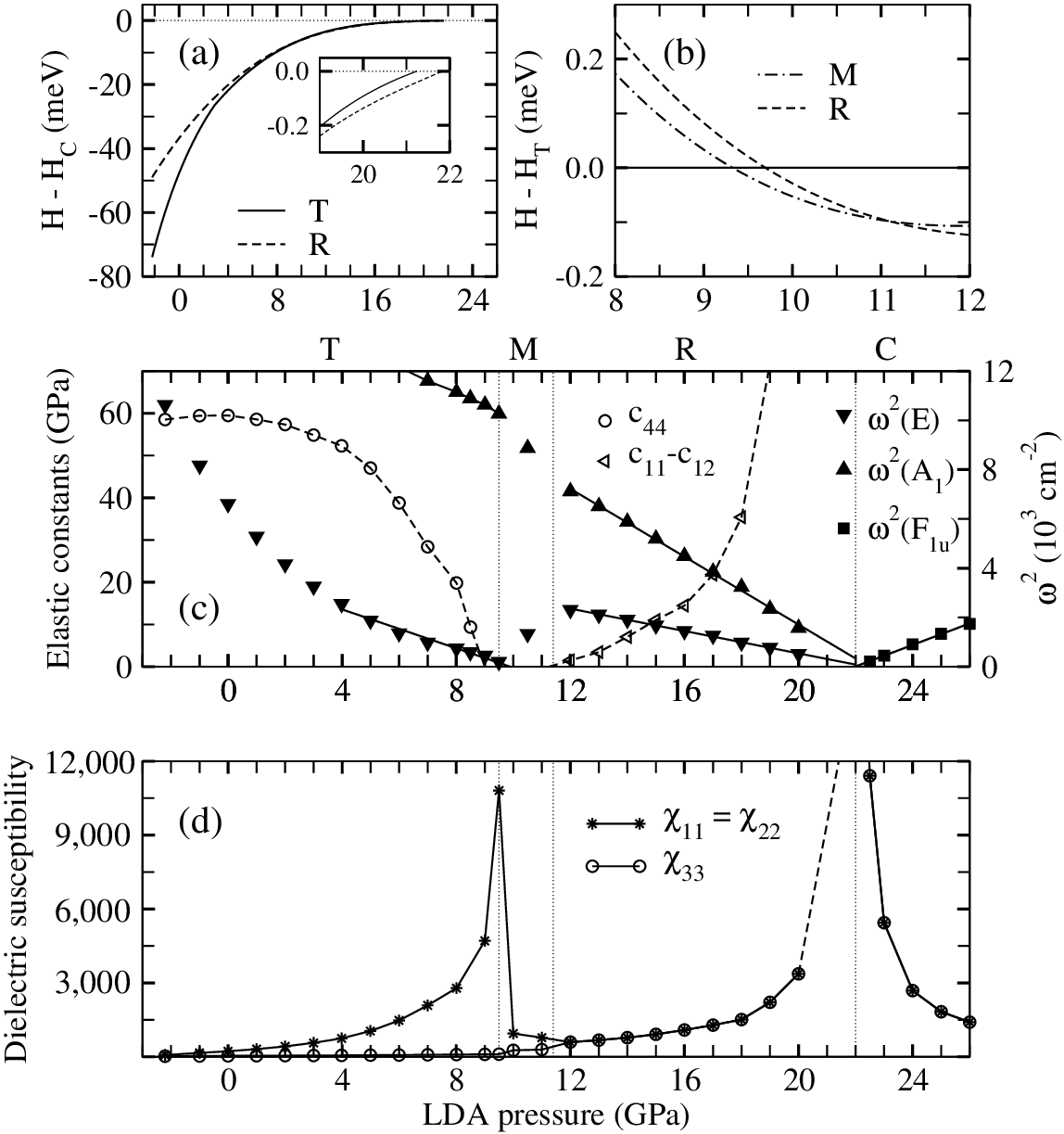}
\end{center}

\caption{\label{fig1} Stability of various phases as a function of pressure. 
(a) Enthalpy difference with respect to the C phase for the T and R phases. 
(b) Enthalpy difference with respect to the T phase for the R and M phases. 
(c) Elastic constant $c_{44}$ and $c_{11} - c_{12}$ respectively for the T and R 
phases, and square of the lowest optical phonon frequencies $\omega^{2}$ for the T 
[$E$(1TO), $A_{1}$(1TO)], R [$E$(1TO), $A_{1}$(1TO)], and C [$F_{1u}$(1TO)] phases. 
(d) Dielectric susceptibility $\chi$. }
\end{figure}

We constrained the symmetry to study four phases: the paraelectric cubic (C) $Pm\bar{3}m$ phase, 
the ferroelectric tetragonal $P4mm$, rhombohedral $R3m$, and monoclinic (M) $Cm$ phases. We find 
that the enthalpy ($H = E + PV$) difference between the T and R phases reduces rapidly with 
pressure and becomes very small when $P > 8$ GPa. The lowest enthalpy corresponds to the 
most stable phase at that pressure. As displayed in Fig. 1(a) and 1(b), there is a T to M phase transition 
around 9.5 GPa ($V = 56.73$ \AA$^{3}$), M to R around 11 GPa ($V = 56.32$ \AA$^{3}$), and R to C 
around 22 GPa ($V = 53.97$ \AA$^{3}$). The M phase for $P = 9.5$ GPa has its spontaneous 
polarization ${\bf P}^{\rm s}$ along the pseudocubic [$uu$1] direction, where $u = 0.481$. The 
discovery of a low symmetry M phase is completely unexpected in a pure compound like PbTiO$_3$, 
and has only been found in complex solid-solutions like PZT \cite{ref18}, PMN-PT \cite{ref19}, 
{\it etc}. 

Because the enthalpy differences among these phases are very small, the above calculated 
pressure-induced phase transition sequence needs to be examined by directly studying
energy derivative properties, such as 
elasticity, phonon frequencies, and dielectricity. The pressure dependence of elastic 
constants $c_{44}$ in the T phase and $c_{11} - c_{12}$ in the R phase is shown in Fig. 1(c). 
Note that the cubic coordinate system is used for all cases, and elastic constants 
$c_{11} = c_{22} = c_{33}$ and $c_{12} = c_{13} = c_{23}$ in the R phase. We find that both 
$c_{44}$ and $c_{11} - c_{12}$ approach zero around 9 and 11.5 GPa, respectively. Negative 
$c_{44}$ and $c_{11} - c_{12}$ mean that the T and R phases are unstable respectively 
against shear and tetragonal shear strain, indicating phase transitions. The estimated phase 
transition pressures from elasticity anomalies agree well with the total energy results. 

In tetragonal $P4mm$ PbTiO$_3$, the $E$(1TO) and $A_{1}$(1TO) modes originate from the triply 
degenerate $F_{1u}$(1TO) mode in the C phase. When the T-to-C phase transition occurs at high 
temperature and ambient pressure \cite{ref2} or high pressure and room temperature \cite{ref3}, both 
the $E$(1TO) and $A_{1}$(1TO) modes soften simultaneously. However, at 0 K we find that pressure 
induces condensation only of the $E$(1TO) mode around $P = 10$ GPa [Fig. 1(c)]. A linear 
combination of the doubly degenerate $E$(1TO) modes gives rise to the transition to the M 
phase. For the R phase, the $E$(1TO) and $A_{1}$(1TO) modes soften simultaneously at 
$P \approx 22$ GPa, going to the $F_{1u}$(1TO) mode in the C phase. This gives rise to the 
R-to-C phase transition. The linearity of the square of mode frequency $\omega^{2}$ vs. pressure 
close to phase transition indicates a Curie-Weiss pressure law. Fig. 1(d) shows that the static 
dielectric susceptibility $\chi_{11}$ has a maximum around 9.5 GPa, indicating a first-order transition; 
while it diverges at $P \approx 22$ GPa, indicating a second-order transition, as can be understood from 
the Lyddane-Sachs-Teller (LST) relation. For the T phase near transition pressure, 
$\chi_{11} = \chi_{22} \propto 1/\omega^{2}[E(1TO)]$, and 
$\chi_{33} \propto 1/\omega^{2}[A_{1}(1TO)]$. 
Thus $\chi_{11}$ increases rapidly around 9.5 GPa, whereas $\chi_{33}$ does not. The same 
analysis can be applied to the R and C phases. The phase 
transition pressures obtained from phonon frequencies and dielectric constants also are consistent 
with the total energy results, and it shows that the computed small energy differences 
between phases are reliable. Our first-principles calcalations neglect quantum fluctuations, which 
could reduce the phase transition pressures \cite{dv2002}.
 
\begin{figure}[tbp]
\begin{center}
\includegraphics[width=\columnwidth]{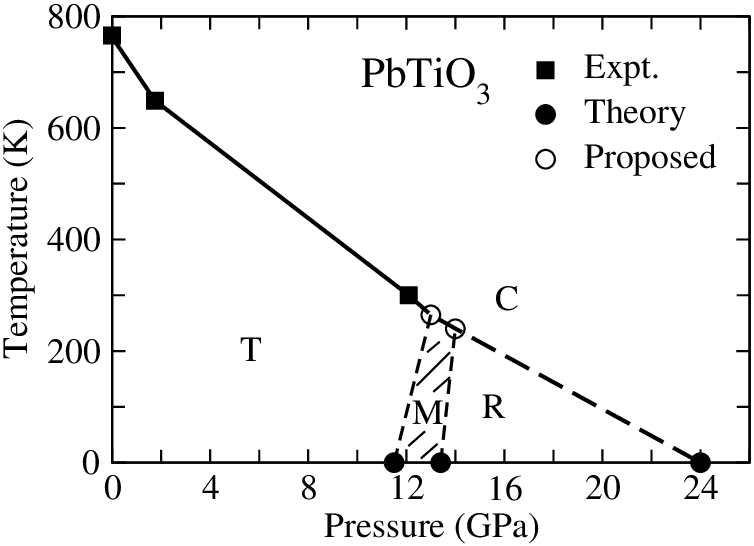}
\end{center}

\caption{\label{fig2} Proposed ($P$, $T$) phase diagram. It 
combines theoretical results (solid circles) at 0 K with experimental data (solid squares) 
at finite temperatures. In this figure we rescaled the theoretical pressure for the 
experimental volume to be zero by shifting. The open circles are our guess for these two tricritical points, 
and an {\it ab\ initio} molecular dynamics simulation is needed to predict them accurately.} 
\end{figure}

We have predicted the pressure-induced phase transitions of PbTiO$_3$ at 0 K. Combining our 
results with experimental finite temperature data, we propose a schematic ($P$, $T$) phase diagram 
(Fig. 2), in which the intermediate M phase separates the T phase at low pressures and the R 
phase at higher pressures. This phase diagram bears remarkable resemblance to the ($x$, $T$) phase 
diagram of PZT \cite{ref18}, where $x$ is the PbTiO$_3$ composition. The narrow M area is the MPB of 
PbTiO$_3$, and it serves as a structure bridging the T and R phases in that its spontaneous 
polarization is located in the ($\bar{1}$10) plane and between the pseudocubic [001] and [111] 
directions, which is very similar to that in PZT. 

\begin{figure}[tbp]
\begin{center}
\includegraphics[width=\columnwidth]{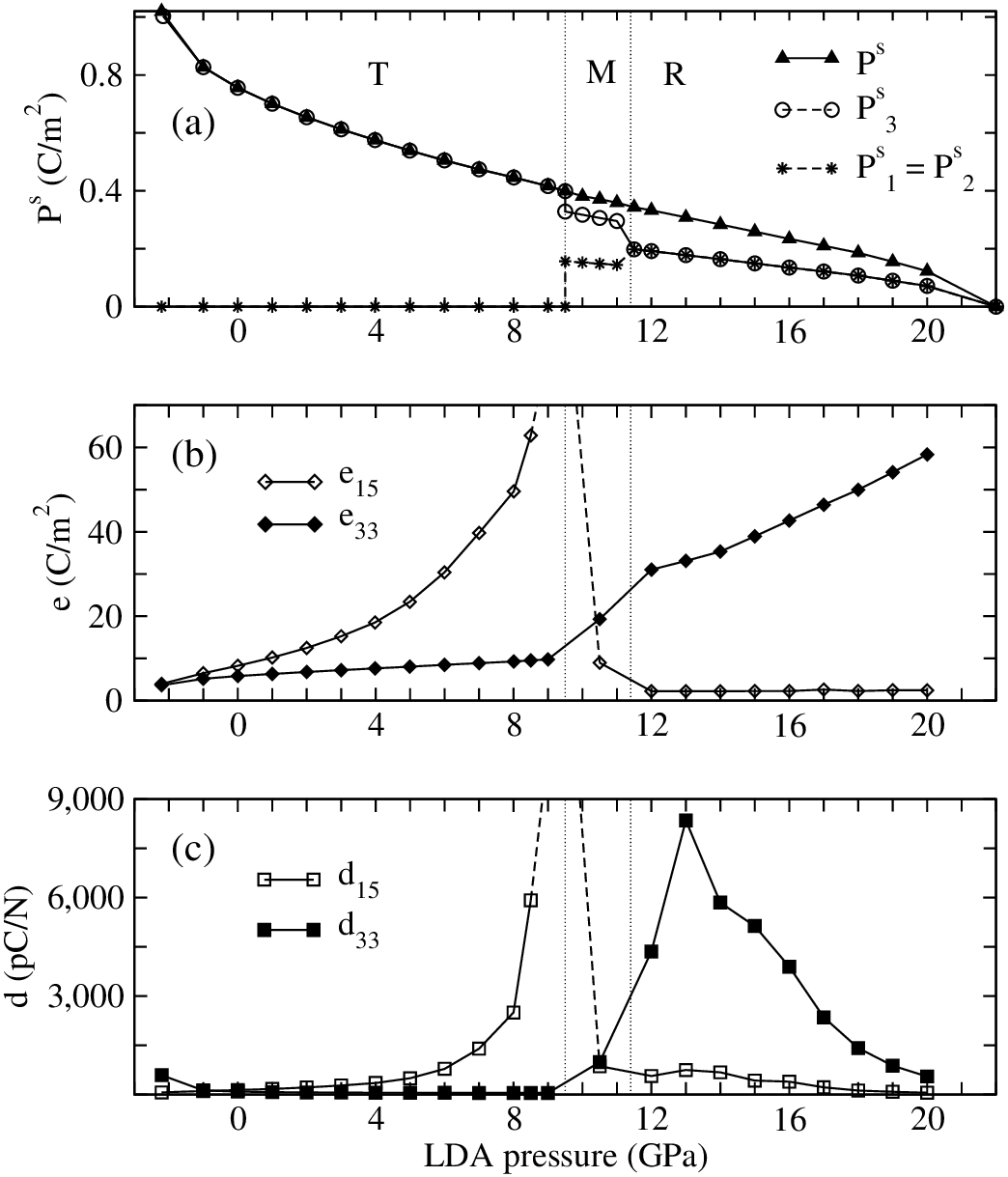}
\end{center}

\caption{\label{fig3} Pressure dependence on piezoelectricity. 
(a) Spontaneous polarization ${\bf P}^{\rm s}$. 
(b) Piezoelectric stress coefficients $e_{15}$ and $e_{33}$. 
(c) Piezoelectric strain coefficients $d_{15}$ and $d_{33}$. }
\end{figure}

Pressure suppresses spontaneous polarization ${\bf P}^{\rm s}$ [Fig. 3(a)] and the $c$-axis 
strain. Although the magnitude of ${\bf P}^{\rm s}$ changes continuously at the T-M transition 
pressure, $P^{\rm s}_{1}$ is discontinuous, indicating a first-order phase transition, consistent 
with the conclusion drawn from dielectric constants. We emphasize that when the T-to-M phase 
transition occurs, ${\bf P}^{\rm s}$ retains about half of the magnitude under zero pressure and 
$c/a = 1.012$. These values are comparable to those of PMN-PT \cite{ref19}, PZN-PT \cite{ref20}, 
and PZT \cite{ref27} in the T phase close to MPB. 

Fig. 3(b) and 3(c) summarize the pressure effect on piezoelectricity of PbTiO$_3$. For the T 
phase, $e_{33}$ increases, whereas the piezoelectric strain coefficient 
$d_{33}$ decreases with pressure. The relatively large $d_{33}$ at $P = -2.2$ GPa 
is due to the LDA overestimation of strain at the experimental volume. Both 
$e_{15}$ and $d_{15}$ increase with pressure, and they rise dramatically for 
$P > 4$ GPa when $c_{44}$ begins to drop quickly [Fig. 1(c)]. The T phase has 
$d_{15} = e_{15}/c_{44}$, so that $d_{15}$ increases even faster than $e_{15}$. 
The pressure-induced large enhancements of $e_{15}$ and $d_{15}$ are a result of the sharp 
reduction of the enthalpy difference between the T and R phases, and both $e_{15}$ and $d_{15}$
diverge at the T-to-M phase transition pressure. For the R phase, both $e_{15}$ and $d_{15}$ are 
rather small, and $e_{33}$ jumps to over 30 C/m$^{2}$, increasing with pressure. $d_{33}$ has 
huge values near the M phase and remains larger than 1500 pN/C for a broad range of pressure. 
The R phase under high pressure has $d_{33} \approx e_{33}/(c_{11} - c_{12})$, and the increase 
of $c_{11} - c_{12}$ with pressure [Fig. 1(c)] causes $d_{33}$ to decrease with pressure,
even though $e_{33}$ rises. Note that $d_{33}$ at 12 GPa is smaller that at 13 GPa, and it could
be due to numerical uncertainties, since small variations of elastic 
constants $c_{14}$ and $c_{15}$ change elastic compliance $s$ 
dramatically when $c_{11}-c_{12}$ is small.  
It is evident that the predicted giant piezoelectric effect of PbTiO$_3$ 
comes from non-collinear polarization rotation since the enhancement of $d_{15}$ in the T phase 
and $d_{33}$ in the R phase do not occur along the spontaneous polarization directions, and the 
M phase acts as a structural bridge between the T and R phases as indicated by pressure inducing 
polarization rotation from [001] to [111] directions. 

We have demonstrated that pressure can induce an MPB in a simple compound PbTiO$_3$, 
which is very similar to the composition-induced MPB in complex solid-solutions, such as 
PZT \cite{ref18}, PMN-PT \cite{ref19}, and PZN-PT \cite{ref20}. It is critical that these 
ferroelectric systems near MPB have distinct phases (different patterns of atomic displacements) 
with very close free energies. Each local minimum is shallow and a 
broad global minimum exists. Low symmetry phases can be stabilized between two high symmetry 
phases if both of them become saddle points, and this is the origin of an MPB. Near an MPB, 
smooth transformation between two high symmetry phases with different polarization directions 
via a low symmetry phase results in strong coupling between internal degree of freedom and strain. 

In PbTiO$_3$, ferroelectricity arises from the competition of short-range repulsions which favor the 
C phase, and Coulomb forces which favor a ferroelectric phase \cite{ref4}. As pressure is increased, 
the short-range repulsions promoting paraelectric stability increase faster than the Coulomb terms 
promoting ferroelectric instability. Therefore pressure favors the paraelectric phase and 
suppresses ferroelectricity. Because the tetragonal well-depth along [001] reduces faster than the 
rhombohedral wells along [111], under a certain pressure the T and R well-depths are equal, and 
the energy surface connecting the T and R phase becomes very flat. Higher order terms in the 
energy expression result in an intermediate M phase \cite{ref28}, which has slightly lower energy than 
both of the T and R phases, leading to a T-to-M phase transition. Further increasing pressure will 
make the R phase more stable than the M phase. For another important ferroelectric, BaTiO$_{3}$, an 
M phase is not expected at low temperature, because the R well-depth is deeper than the T 
well-depth so that pressure can not induce an M phase before both well-depths disappear. 

Pressure-tuning of the structural and piezoelectric properties of PbTiO$_3$ promises to be an 
intriguing avenue for experimental study. PbTiO$_3$ under pressure should be the simplest possible 
system to study the basic physics of the MPB and piezoelectric enhancement through polarization 
rotation. The predicted monoclinic $Cm$ phase (without strain and external electric field) and the 
MPB accompanied by giant piezoelectric effects in simple perovskite compounds have never yet 
been reported. Very recently, Rouquette {\it et al.} reported experimental evidence of polarization
rotation induced by pressure in PZT \cite{ref29,ref30,ref31}.
We hope our predictions will stimulate both fundamental and technological 
interests to investigate ferroelectrics under hydrostatic or uniaxial pressure, and stress and 
pressure as means to enhance piezoelectricity. 

We thank fruitful discussions with N. Choudhury and E. J. Walter. 
Computations were done on the Center for Piezoelectrics by Design. This work was supported by 
the Office of Naval Research under ONR Grants No. N00014-02-1-0506.


\begin{references}

\bibitem{ref2} 
G. Burns and B. A. Scott, 
Phys. Rev. Lett. {\bf 25}, 167 (1973). 

\bibitem{ref3}  
J. A. Sanjurjo, E. L\'{o}pez-Cruz, and G. Burns, 
Phys. Rev. B {\bf 28}, 7260 (1983). 

\bibitem{ref9}
G. Shirane, S. Hoshino, and K. Suzuki, 
Phys. Rev. {\bf 80}, 1105 (1950). 

\bibitem{ref10}
M. E. Lines and A. M. Glass, 
{\it Principles and Applications of Ferroelectrics and Related Materials} 
(Clarendon Press, Oxford, 1977). 

\bibitem{ref11}
K. Uchino, 
{\it Piezoelectric acutuators and ultrasonic motors} 
(Kluwer Academic, Boston, 1996).


\bibitem{ref1} 
{\it Chemical Abstracts} reports 9924 publications on PbTiO$_3$ since its discovery. 

\bibitem{ref4} 
R. E. Cohen, 
Nature (London) {\bf 358}, 136 (1992). 

\bibitem{ref5} 
A. Garc\'{i}a and D. Vanderbilt, 
Phys. Rev. B {\bf 54}, 3817 (1996). 

\bibitem{ref6} 
U. V. Waghmare and K. M. Rabe, 
Phys. Rev. B {\bf 55}, 6161 (1997). 

\bibitem{ref7}
G. S\'{a}ghi-Szab\'{o}, R. E. Cohen, and H. Krakauer, 
Phys. Rev. Lett. {\bf 80}, 4321 (1998).
 
\bibitem{ref8}
Z. Wu, G. S\'{a}ghi-Szab\'{o}, R. E. Cohen, and H. Krakauer, 
Phys. Rev. Lett. {\bf 94}, 069901 (2005). 

\bibitem{ref12} 
R. Ramirez, M. F. Lapena, and J. A. Gonzalo, 
Phys. Rev. B {\bf 42}, 2604 (1990). 

\bibitem{ref13} 
S. E. Park and T. R. Shrout, 
J. of Appl. Phys. {\bf 82}, 1804 (1997). 

\bibitem{ref14} 
H. Fu and R. E. Cohen, 
Nature (London) {\bf 403}, 281 (2000). 

\bibitem{ref15} 
B. Noheda, {\it et al.},
Phys. Rev. Lett. {\bf 86}, 3891 (2001). 

\bibitem{ref16} 
L. Bellaiche, A. Garc\'{i}a, and D. Vanderbilt, 
Phys. Rev. Lett. {\bf 84}, 5427 (2000). 

\bibitem{ref17} 
Z. Wu and H. Krakauer, 
Phys. Rev. B {\bf 68}, 014112 (2003). 

\bibitem{ref18} 
B. Noheda, {\it et al.},
Appl. Phys. Lett. {\bf 74}, 2059 (1999). 

\bibitem{ref19} 
B. Noheda, D. E. Cox, G. Shirane, J. Gao, and Z.-G. Ye, 
Phys. Rev. B {\bf 66}, 054104 (2002). 

\bibitem{ref20} 
D. La-Orauttapong, {\it et al.},
Phys. Rev. B {\bf 65}, 144101 (2002). 

\bibitem{ref21} 
X. Gonze, {\it et al.},
Comput. Mater. Sci. {\bf 25}, 478 (2002).

\bibitem{ref22} 
A. M. Rappe, K. M. Rabe, E. Kaxiras, and J. D. Joannopoulos, 
Phys. Rev. B {\bf 41}, 1227 (1990). 

\bibitem{ref23} 
D. J. Singh, 
{\it Planewaves, Pseudopotentials, and the LAPW Method} 
(Kluwer Academic Publishers, Boston, 1994). 

\bibitem{ref24} 
Z. Wu, R. E. Cohen, and D. J. Singh, 
Phys. Rev. B {\bf 70}, 104112 (2004). 

\bibitem{ref25}
S. Baroni, S. de Gironcoli, A. Dal Corso, and P. Giannozzi, 
Rev. Mod. Phys. {\bf 73}, 515 (2001). 

\bibitem{ref26} 
D. R. Hamann, X. Wu, K. M. Rabe, and D. Vanderbilt, 
Phys. Rev. B {\bf 71}, 035117 (2005). 

\bibitem{dv2002}
J. \'{I}\~{n}iguezD. Vanderbilt,
Phys. Rev. Lett. {\bf 89}, 115503 (2002).

\bibitem{ref27} 
B. Noheda, {\it et al.},
Phys. Rev. B {\bf 63}, 014103 (2001). 

\bibitem{ref28} 
D. Vanderbilt and M. H. Cohen, 
Phys. Rev. B {\bf 63}, 094108 (2001). 

\bibitem{ref29}
J. Rouquette, {\it et al.},
Phys. Rev. B {\bf 71}, 024112 (2005). 

\bibitem{ref30}
J. Rouquette, {\it et al.},
Phys. Rev. B {\bf 70}, 014108 (2004). 

\bibitem{ref31}
J. Rouquette, {\it et al.},
Phys. Rev B 65, 214102 (2002))

\end{references}
\end{document}